\documentclass[%
 reprint,
superscriptaddress,
 amsmath,amssymb,
 aps,
prl
]{revtex4-2}

\usepackage{graphicx}
\usepackage{dcolumn}
\usepackage{bm}

\usepackage{xcolor}

\begin{document}

\preprint{APS/123-QED}

\title{Spontaneous PT-symmetry breaking transitions under the influence of noise in an optomechanical system}

\author{A. R. Mukhamedyanov}
\affiliation{Moscow Institute of Physics and Technology, 141700, 9 Institutskiy pereulok, Moscow, Russia}

\author{E. S. Andrianov}
\affiliation{Moscow Institute of Physics and Technology, 141700, 9 Institutskiy pereulok, Moscow, Russia}
\affiliation{Dukhov Research Institute of Automatics (VNIIA), 127055, 22 Sushchevskaya, Moscow, Russia}
\affiliation{Institute for Theoretical and Applied Electromagnetics, 125412, 13 Izhorskaya, Moscow, Russia}

\author{A. A. Zyablovsky}
\email{zyablovskiy@mail.ru}
\affiliation{Moscow Institute of Physics and Technology, 141700, 9 Institutskiy pereulok, Moscow, Russia}
\affiliation{Dukhov Research Institute of Automatics (VNIIA), 127055, 22 Sushchevskaya, Moscow, Russia}
\affiliation{Institute for Theoretical and Applied Electromagnetics, 125412, 13 Izhorskaya, Moscow, Russia}

\date{\today}

\begin{abstract}
Spontaneous symmetry breaking is a phenomenon of an alteration of a state symmetry without a change in the system symmetry. A transition from a state with unbroken symmetry to a state with broken symmetry leads to a qualitative change in system properties, and respective states are called as symmetric and non-symmetric phases of the system. Usually, the change in the system parameters is necessary for the transition between symmetric and non-symmetric phases. In this letter, we consider the behavior of an optomechanical system with an exceptional point under the influence of noise. We demonstrate that below the exceptional point, PT-symmetric and non-PT-symmetric phases can coexist. In this regime, a noise can lead to random transitions between the symmetric and non-symmetric phases. These transitions are accompanied by the abrupt changes in the intensities of photon and phonon modes. The predicted phenomenon opens up the possibility of studying the kinetics of phase transitions caused by spontaneous breaking of PT symmetry.
\end{abstract}

\maketitle

\textit{Introduction.} The phenomenon of spontaneous symmetry breaking is a one of central concepts in many areas of modern physics ranging from high-energy \cite{8} and solid state \cite{2} physics to optics \cite{3, 7, 6, 5}. Spontaneous symmetry breaking can result in the emergence of new features and behaviors that were not initially present in the system, making it a topic of great interest. This effect often occurs due to interaction between various subsystems, for example, between the molecules in a crystal lattice \cite{2}, and often leads to form a certain collective state in the system. In early works, spontaneous symmetry breaking has been studied in the macroscopic systems, in which a number of the degrees of freedom is much greater than one \cite{2}. In recent years, spontaneous symmetry breaking began to be studied in systems with only several degrees of freedom. An example of such systems is parity-time (PT)-symmetric ones \cite{4, 3, 7, 6, 5}, in which a combination of gain and loss leads to an existence of regions in space of system parameters with PT-symmetric and non-PT-symmetric eigenstates \cite{4, 3, 7, 6, 5}. The change in the magnitudes of gain/loss leads to transitions between these states that are a manifestation of spontaneous breaking of PT-symmetry \cite{4, 3, 7, 6, 5}. These transitions occur at exceptional points \cite{4, 3, 7, 6, 5}, in which two or mode eigenstates coalesce and changes their symmetry.

Exceptional points (EPs) take place in various types of non-Hermitian systems: optical \cite{10,7,6,5,11,9,38} and magnonic \cite{12,47,15,18,16,17,19} systems with gain and loss, optomechanical \cite{20,21,23,22,44,45,46} and laser \cite{24, 29, 27, 14, 13, 28, 30, 25, 26} systems, etc. In these systems, the passing through EPs leads to spontaneous breaking of symmetry in the eigenstates \cite{4, 3, 7, 6, 5}. 
The passing through EPs occurs at an alteration of system parameters. There are areas of the system parameters, in which the eigenstates are symmetric, and are non-symmetric. The transitions between the areas are accompanied by changes in the system behavior, which is associated with phase transitions at EP \cite{3, 6, 5}. Such transitions find applications for creation of new generations of lasers \cite{24, 29, 27, 14, 13, 28, 25}, sensors \cite{34, 20, 31, 32, 36, 33}, etc.

In this letter, we demonstrate for the first time that noise are also parameters that can alter symmetry of the system state. On an example of an optomechanical system with an exceptional point, we show that there may be a parameters range in which PT-symmetric and non-PT-symmetric states coexist. In the symmetric state the intensities of optical and phonon oscillations are equal to each other. In non-symmetric state, the intensity of phonon oscillations is greater than the intensity of the optical one. Under the influence of noise, the system evolves between such states, switching from the PT-symmetric phase to the non-PT-symmetric one and vice-versa. These transitions under the influence of noise are observed below the exceptional point. The predicted phenomenon sheds light on the kinetics of phase transitions caused by spontaneous breaking of PT-symmetry.

\textit{Model.} We consider an optomechanical system consists of two optical modes interacting with each other via a phonon mode. We suppose that the frequency difference between the optical modes is equal to the frequency of the phonon mode. The external electromagnetic wave excites an optical mode with a higher frequency.
To describe the considered system we use the optomechanical Hamiltonian \cite{1}:

\begin{equation}
\begin{gathered}
  \hat H = \hbar {\omega _1}\hat a_1^\dag {{\hat a}_1} + \hbar {\omega _2}\hat a_2^\dag {{\hat a}_2} + \hbar {\omega _b}{{\hat b}^\dag }\hat b +  \hfill \\
  \hbar \,g\left( {\hat a_1^\dag {{\hat a}_2}\hat b + {{\hat a}_1}\hat a_2^\dag {{\hat b}^\dag }} \right) + \hbar \Omega \left( {\hat a_1^\dag {e^{ - i\omega t}} + {{\hat a}_1}{e^{i\omega t}}} \right) \hfill \\ 
\end{gathered}
\label{eq:1}
\end{equation}

Here ${\hat a_{1,2}}$ and $\hat a_{1,2}^\dag $ are the annihilation and creation bosonic operators for the first and the second optical modes, respectively. $\hat b$ and ${\hat b^\dag }$ are the annihilation and creation operators of the phonon mode. ${\omega _{1,2}}$ is frequencies of the optical modes. ${\omega _b}$ is a frequency of the phonon mode. $g$ is a coupling strength between the optical modes and the phonon mode (Frohlich constant). $\Omega $ is an amplitude of the external wave. $\omega $ is a frequency of the external wave.
Within the framework of Heisenberg-Langevin approach \cite{39}, we derive the equations for the averages of amplitudes of the optical modes and phonon mode that have the form \cite{37} 

\begin{equation}
\frac{{d{a_1}}}{{dt}} = \left( { - {\gamma _1} - i{\omega _1}} \right){a_1} - ig{a_2}b - i\Omega {e^{ - i\,\omega \,t}}
\label{eq:2}
\end{equation}

\begin{equation}
\frac{{d{a_2}}}{{dt}} = \left( { - {\gamma _2} - i{\omega _2}} \right){a_2} - i{g}{a_1}{b^*}
\label{eq:3}
\end{equation}

\begin{equation}
\frac{{db}}{{dt}} = \left( { - {\gamma _b} - i{\omega _b}} \right)b - i{g}{a_1}a_2^* + \xi \left( t \right)
\label{eq:4}
\end{equation}

Here ${a_{1,2}} = \left\langle {{{\hat a}_{1,2}}} \right\rangle $, $b = \left\langle {\hat b} \right\rangle $ are the averages of amplitudes of the optical modes and phonon mode, respectively. ${\gamma _{1,2}}$, ${\gamma _b}$ are the relaxation rate of the respective quantities. $\xi \left( t \right)$ is a thermal noise acting on the phonon mode, the correlation function of which is proportional to ${\gamma _b}\bar n$, where $\bar n$ is an average number of thermal phonon in the system.
We consider that the system temperature is much less than the frequencies of the optical modes ($kT <  < \hbar {\omega _{\,1,2}}$) and, therefore, we neglect a noise acting on the optical modes \cite{37}. We also consider that the relaxation rates of the second optical mode and the phonon mode are the same (${\gamma _2} = {\gamma _b} = \gamma $).

\textit{Stationary states.} To start, we study the system behavior without taking into account noise. We are looking for a stationary solution of Eqns.~(\ref{eq:2})-(\ref{eq:4}) in the form ${a_1} = {a_{1st}}\exp \left( { - i\omega \,t} \right)$, ${a_2} = {a_{2st}}\exp \left( { - i\left( {\omega  - \delta \omega } \right)t} \right)$, $b = {b_{st}}\exp \left( { - i\delta \omega t} \right)$, where ${a_{1st}}$, ${a_{2st}}$, ${b_{st}}$ do not depend on time and $\delta \omega$ is a frequency of generated phonons, which is determined from Eqns.~(\ref{eq:2})-(\ref{eq:4}) \cite{37}.

In the considered system, there is \textit{a zero solution}, for which ${a_{2st}} = {b_{st}} = 0$, ${a_{1st}} =  - i\,\Omega /\left( {i\delta {\omega _{\,1}} + {\gamma _1}} \right)$. Small deviations of $a_2$ and ${b^*}$ from the zero solution are determined by the following equations:

\begin{equation}
\frac{d}{{dt}}\left( {\begin{array}{*{20}{c}}
  {\delta {a_2}} \\ 
  {\delta b} 
\end{array}} \right) = \left( {\begin{array}{*{20}{c}}
  { - \gamma  - i\delta {\omega _2}}&{ - \kappa } \\ 
  { - {\kappa ^*}}&{ - \gamma  + i\,{\omega _b}} 
\end{array}} \right)\left( {\begin{array}{*{20}{c}}
  {\delta {a_2}} \\ 
  {\delta b} 
\end{array}} \right)
\label{eq:5}
\end{equation}
where $\kappa  = \frac{{g\,\Omega }}{{{\gamma _1} + i\delta {\omega _{\,1}}}}$ and $\delta {\omega _{\,1,2}} = {\omega _{\,1,2}} - \omega$. The matrix on the right part of Eq. (5) can be represented as the identity matrix multiplied by the coefficient and the anti-PT-symmetric matrix (an anti-PT-symmetric matrix differs from PT-symmetric matrix by multiplication by the imaginary unit \cite{40,41,42,43})

\begin{equation}
\begin{gathered}
  \left( {\begin{array}{*{20}{c}}
  { - \gamma  - i\delta {\omega _2}}&{ - \kappa } \\ 
  { - {\kappa ^*}}&{ - \gamma  + i\,{\omega _b}} 
\end{array}} \right) =  \hfill \\
  \left( { - \gamma  - i\frac{{\delta {\omega _2} - {\omega _b}}}{2}} \right)\left( {\begin{array}{*{20}{c}}
  1&0 \\ 
  0&1 
\end{array}} \right) + \left( {\begin{array}{*{20}{c}}
  { - i\Delta }&{ - \kappa } \\ 
  { - {\kappa ^*}}&{i\,\Delta } 
\end{array}} \right) \hfill \\ 
\end{gathered}
\label{eq:6}
\end{equation}
Here $\Delta  = \left( {\delta {\omega _2} + {\omega _b}} \right)/2$.

Eigenvalues of Eq.~(\ref{eq:5}) are given as

\begin{equation}
{\lambda _ \pm } =  - \gamma  - i\frac{{\delta {\omega _2} - {\omega _b}}}{2} \pm \sqrt {{{\left| \kappa  \right|}^2} - {\Delta ^2}}
\label{eq:7}
\end{equation}

and the eigenvectors are

\begin{equation}
{{\mathbf{e}}_ \pm } = \left( {\begin{array}{*{20}{c}}
  {\frac{{i\Delta  \mp \sqrt {{{\left| \kappa  \right|}^2} - {\Delta ^2}} }}{{{\kappa ^*}}}} \\ 
  1 
\end{array}} \right)
\label{eq:8}
\end{equation}

It is seen that at $\Omega  = {\Omega _{EP}} =  \pm \frac{{\sqrt {\gamma _1^2 + \delta \omega _{\,1}^2} }}{g}\frac{{\left| {\delta {\omega _2} + {\omega _b}} \right|}}{2}$ (i.e. when $\left| \kappa  \right| = \left| \Delta  \right|$) there is an exceptional point, for which the eigenvalues are equal to each other and the eigenvectors coincide. When $\Omega  < {\Omega _{EP}}$, the eigenvectors are non-PT-symmetric ($\left| {\delta {a_2}} \right| \ne \left| {\delta b} \right|$) \cite{38}. At the same time, when $\Omega  > {\Omega _{EP}}$, the eigenvectors are PT-symmetric ($\left| {\delta {a_2}} \right| = \left| {\delta b} \right|$) \cite{38}. Namely, if we swap the first and second components of a given eigenvector, and after that take its complex conjugate (PT transformation: ${\left( {{{\mathbf{e}}_{1,2}}} \right)_ \pm } \to \left( {{{\mathbf{e}}_{2,1}}} \right)_ \pm ^*$) \cite{38}, then the eigenvector will change only by the common coefficient. This coefficient is equal to the eigenvalue of the PT transformation. Thus, the transition through the exceptional point is equivalent to a spontaneous breaking of PT-symmetry \cite{38}. Note that since the matrix of the anti-PT-symmetric system differs from the matrix of the PT-symmetric system only by a factor, their eigenvectors coincide. Therefore, the transition through the exceptional point leads to breaking PT-symmetry of the eigenvectors.

The real parts of both eigenvalues \ref{eq:7} are less than zero when $\Omega  < {\Omega _{th}} = \frac{1}{{g}}\sqrt {{{\left( {\delta {\omega _{\,1}}\,\gamma  + {\gamma _1}\,\Delta } \right)}^2} + {{\left( {{\gamma _1}\gamma  - \delta {\omega _{\,1}}\,\Delta } \right)}^2}} $, where $\Delta  = \left( {\delta {\omega _2} + {\omega _b}} \right)/2$. Therefore, the zero solution is stable when $\Omega  < {\Omega _{th}}$ \cite{37}.

Besides the zero solution, there is \textit{a nonzero solution} of Eqns.~(\ref{eq:2})-(\ref{eq:4}) without noise, for which the stationary values of the amplitudes of the second optical modes and the phonon mode, are given as

\begin{equation}
{\left| {{a_{2 \pm }}} \right|^2} =  \pm \frac{1}{{g}}\sqrt {{{\left| \Omega  \right|}^2} - \Omega _{ex}^2}  + \frac{{\delta {\omega _{\,1}}\,\Delta  - {\gamma _1}\,\gamma }}{{{{g}^2}}}
\label{eq:9}
\end{equation}

\begin{equation}
{\left| {{b_ \pm }} \right|^2} =  \pm \frac{1}{{g}}\sqrt {{{\left| \Omega  \right|}^2} - \Omega _{ex}^2}  + \frac{{\delta {\omega _{\,1}}\,\Delta  - {\gamma _1}\,\gamma }}{{{{g}^2}}}
\label{eq:10}
\end{equation}
where ${\Omega _{ex}} = \left| {\frac{{\delta {\omega _{\,1}}\,\gamma  + {\gamma _1}\,\Delta }}{g}} \right|$ \cite{37}. This solution describes the laser generation in the system. The frequency of generated phonons is $\delta \omega  = \frac{{{\omega _b}\,{\gamma _2} - \delta {\omega _2}{\gamma _b}}}{{{\gamma _2} + {\gamma _b}}}$.

It is seen that the nonzero solution is symmetric, i.e., ${\left| {{a_{2st}}} \right|^2} = {\left| {{b_{st}}} \right|^2}$. This equality holds when ${\gamma _2} = {\gamma _b}$ and does not depend on the ratio between the frequencies of the second optical mode and the phonon mode. In addition, the symmetry of the nonzero solution does not depend on the amplitude of external wave whether it is less or greater than one corresponding to the exceptional point.

The independence of the ratio of the intensities of the second optical mode and phonon mode from the frequencies of the modes is due to the fact that the generation frequency, $\delta \omega$, self-adjusts in such a way that $\left| {{a_{2st}}} \right| = \left| {{b_{st}}} \right|$. Fulfilment of this equality ensures maximum interaction between the optical and phonon modes, which promotes laser generation. Therefore, the transition to laser generation leads to the appearance of symmetry in the system, which is clearly not present in the Eqs.~(\ref{eq:2})-(\ref{eq:4}) and in the  Hamiltonian~(\ref{eq:1}). The independence of the symmetry of the nonzero solution from the mode frequencies is extremely important, because the parameter region in which the eigenvectors for the zero solution are PT-symmetric depends on the mode frequencies. This allows tuning the mode frequencies to achieve the existence of a parameter region in which the eigenvectors for the zero solution are non-symmetric, and the nonzero solution is symmetric.

From a linear analysis of stability for the nonzero solution, we obtain that if $\delta {\omega _{\,1}}\,\Delta  > {\gamma _1}\gamma $, the nonzero solution is stable when $\Omega  \geqslant {\Omega _{ex}}$. If $\delta {\omega _{\,1}}\,\Delta  < {\gamma _1}\,\gamma $, the nonzero solution is stable when $\Omega  \geqslant {\Omega _{th}}$ \cite{37}. Thus, when $\delta {\omega _{\,1}}\,\Delta  > {\gamma _1}\,\gamma $, there is an area of the coupling strength ${\Omega _{ex}} < \Omega  < {\Omega _{th}}$, in which both the zero solution and the nonzero solution are stable ($\Omega_{ex} \le  \Omega_{th}$ for all values of the parameters). That is, with such parameters there exists a bistable area. 

By choosing the frequencies of optical and phonon modes, we can achieve the fulfillment of inequalities ${\Omega _{ex}} < {\Omega _{EP}} < {\Omega _{th}}$. In this case, with ${\Omega _{ex}} < \Omega  < {\Omega _{EP}}$, there coexists the zero solution whose eigenvectors are non-symmetric, and the nonzero solution that is symmetric. That is, it is possible to achieve the coexistence of solutions with different symmetries. Further, we study the system dynamics in this parameter region under influence of the noise.

\begin{figure}[htbp]
\centering\includegraphics[width=\linewidth]{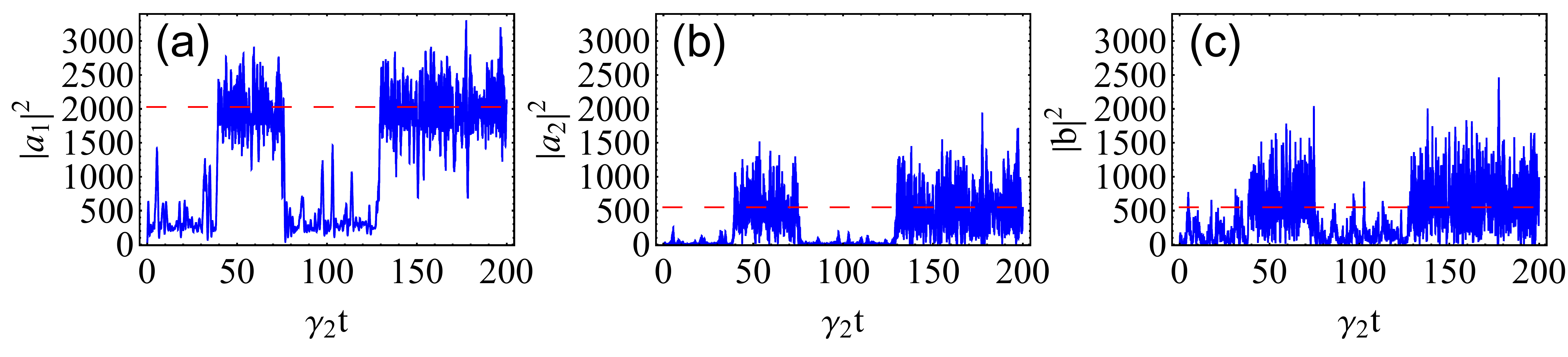}
\caption{Time dependence of the intensity of the first optical mode (a), the second optical mode (b) and the phonon mode (c). The horizontal dashed lines show ${\left| {{a_{1st}}} \right|^2}$ (a), ${\left| {{a_{2st}}} \right|^2}$ (b) and ${\left| {{b_{st}}} \right|^2}$ (c) for the nonzero solution. ${\Omega _{ex}} < \Omega  < {\Omega _{EP}}$. Here ${\gamma _1} = {\gamma _2} = {\gamma _b} = 2 \cdot {10^{ - 4}}{\omega _0}$ , $\delta {\omega _{\,1}} = {10^{ - 3}}{\omega _0}$ , $\delta {\omega _2} = 5 \cdot {10^{ - 3}}{\omega _0}$ , ${\omega _b} = 4 \cdot {10^{ - 3}}{\omega _0}$ , $g = {10^{ - 4}}{\omega _0}$ , $\Omega  = 1.5 \cdot {10^{ - 2}}{\omega _0}$ and $n = {10^2}$.}
\label{fig:1}
\end{figure}

\textit{A transition between symmetric and non-symmetric phases under the influence of noise.} We study the system behavior in the bistable region when $\Omega  < {\Omega _{EP}}$. To study the dynamics of the system with noise, we numerically simulate Eqns.~(\ref{eq:2})-(\ref{eq:4}). In the previous section, we show that when $\Omega  < {\Omega _{EP}}$, the eigenvectors of Eq.~(\ref{eq:5}) for small deviation from the zero solution are non-symmetric and the nonzero solution is symmetric. In the bistable region, depending on the initial conditions, the system evolves into one of the stable states [Figure~\ref{fig:1}]. The presence of noise leads to fluctuations of the system near stable states, and can also lead to transitions of the system from one state to another [Figure~\ref{fig:1}]. Such transitions are random, and their probability increases with the amplitude of the noise.

\begin{figure}[htbp]
\centering\includegraphics[width=0.7\linewidth]{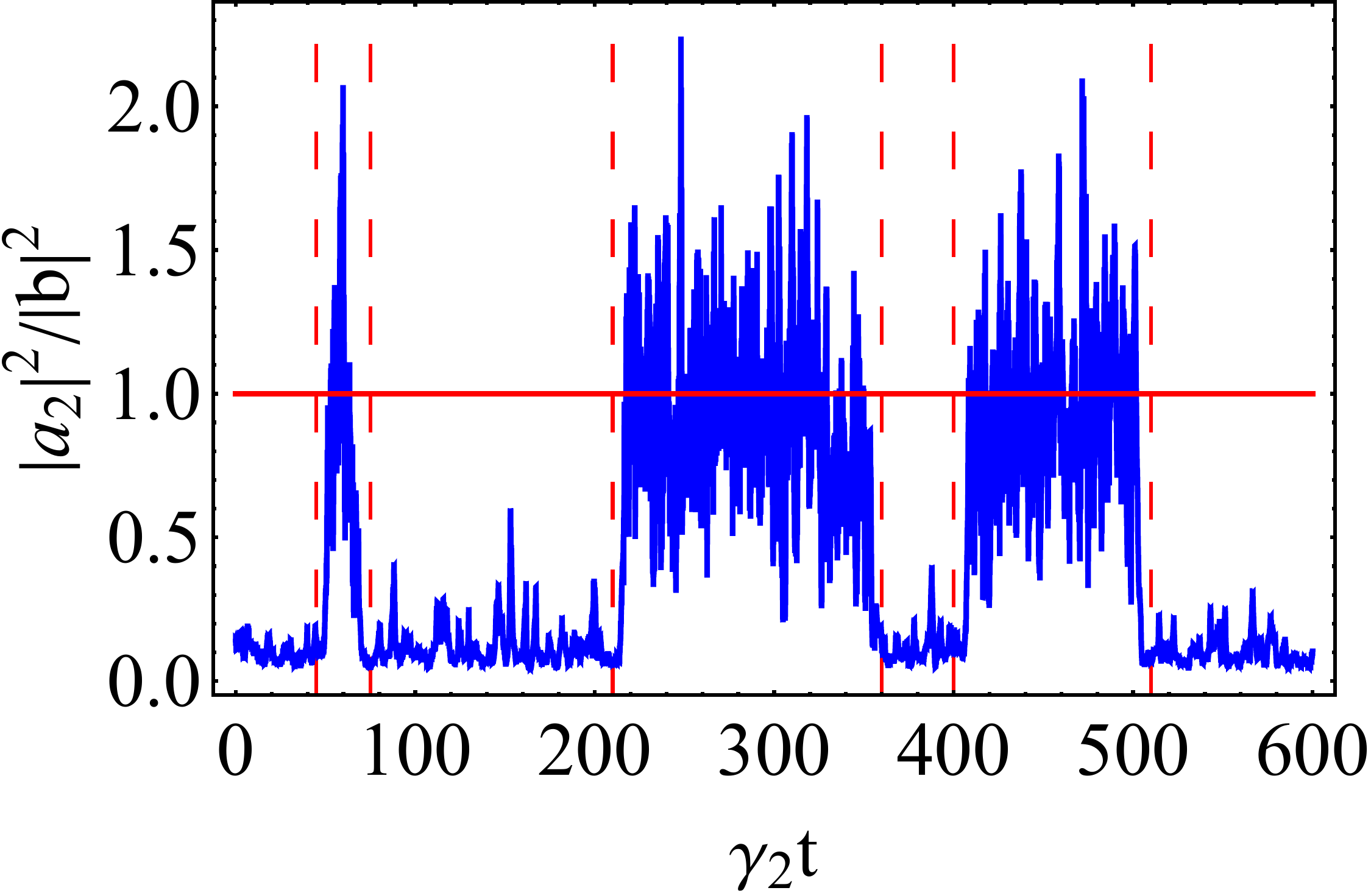}
\caption{Time dependence of the ratio of the intensity of the second optical mode to the intensity of the phonon mode. ${\Omega _{ex}} < \Omega  < {\Omega _{EP}}$ and ${\gamma _2} = {\gamma _b}$. The dashed lines separate the time intervals when the system evolves near the PT-symmetric state (${\left| {{a_2}} \right|^2}/{\left| b \right|^2} = 1$) and when it evolves near the non-PT-symmetric state (${\left| {{a_2}} \right|^2}/{\left| b \right|^2} < < 1$). All parameters are as in Figure~\ref{fig:1}.}
\label{fig:2}
\end{figure}

In the case when the system evolves near the nonzero solution, noise leads to relatively small fluctuations in intensity. As a result, the averaged intensities of the first and second optical modes, $\left\langle {{{\left| {{a_{1,2}}} \right|}^2}} \right\rangle $, and the phonon mode, $\left\langle {{{\left| b \right|}^2}} \right\rangle $, are close to the values ${\left| {{a_{1st}}} \right|^2}$, ${\left| {{a_{2st}}} \right|^2}$ and ${\left| {{b_{st}}} \right|^2}$ corresponding to the nonzero solution. Since ${\left| {{a_{2st}}} \right|^2} = {\left| {{b_{st}}} \right|^2}$ then $\left\langle {{{\left| {{a_2}} \right|}^2}} \right\rangle  = \left\langle {{{\left| b \right|}^2}} \right\rangle$ [Figure~\ref{fig:2}]. Such a regime is a symmetric phase of the system.

In the case where the system evolves near the zero state, the noise completely determines the intensities of the second optical and phonon modes. The noise excites oscillations near the zero state, the subsequent dynamics of which are determined by Eq.~(\ref{eq:5}). Since the noise acts on the phonon mode, the eigenvector \ref{eq:8} in which the phonon contribution prevails ($\left| {\delta b} \right| > \left| {\delta {a_2}} \right|$) is mainly excited. As a result, in this case the system is in a non-symmetric phase [Figure~\ref{fig:2}].

In addition to fluctuations near steady states, noise causes random transitions between states. For this reason, the system randomly switches from the non-symmetric phase (the state with lower intensity) to the symmetric phase (the state with higher intensity) and back [Figure~\ref{fig:2}]. These switches induced by the noise are transitions with spontaneous symmetry breaking.
 
Above we consider the case when $\Omega  < {\Omega _{EP}}$. When $\Omega  > {\Omega _{EP}}$, the eigenvectors~(\ref{eq:8}) are symmetric ($\left| {\delta {a_2}} \right| = \left| {\delta b} \right|$) and the nonzero solution is symmetric too. As a result, the transitions from the zero solution to the nonzero solution do not lead to changes in symmetry of the system state.

\textit{Discussion.} The experimental observation of the spontaneous symmetry breaking transition caused by the noise requires measuring the intensities of optical and phonon modes. The transition can be detected by sharp change in the intensities of the modes. To prove that the transition is accompanied by the change in symmetry, it is necessary to find the temporal dependence of the ratio of the intensities of the optical and phonon modes. In case of phonon lasers, which serve as sources of the sound waves, the measurement of the phonon mode intensity is not difficult. However, in optomechanical systems, where the focus is on photon generation (e.g., Brillouin laser), such measurement can be challenging.

Indirect manifestation of the spontaneous symmetry breaking in the optomechanical system is a sharp dependence of the intensity of the second optical mode on the external wave amplitude near the lasing threshold. Indeed, in the non-symmetric phase, the intensity of the second optical mode is much less than the one of the phonon mode. That is, the intensity of the second optical mode is suppressed in the non-symmetric phase. While, the intensity of the secondoptical mode is not suppressed in the symmetric phase. As a result, the transitions between states are accompanied by a large relative change in the intensity of the second optical mode. Below the lasing threshold, when $\Omega  < {\Omega _{ex}}$, the system evolves near the zero solution (the non-symmetric phase). Above the lasing threshold, the system evolves near a nonzero solution (the non-symmetric phase). As a result, an increase in the pump wave amplitude from $\Omega  < {\Omega _{ex}}$ to $\Omega  > {\Omega _{th}}$ leads to a stronger relative increase in the intensity of the second optical mode.

\textit{Conclusion.} We demonstrate that in an optomechanical system, PT-symmetric and non-PT-symmetric phases can coexist. In the symmetric phase the intensities of optical and phonon modes are equal to each other. While in the non-symmetric phase, the intensity of phonon mode is greater than the intensity of the optical one. The presence of noise in such a system leads to random transitions between the symmetric and non-symmetric phases. These transitions induced by the noise are transitions with spontaneous PT-symmetry breaking. The proposed system is a convenient object for studying kinetics of phase transitions caused by spontaneous PT-symmetry breaking.

\section*{Acknowledgments}
The study was financially supported by a Grant from Russian Science Foundation (project No. 20-72-10057). A.R.M. and E.S.A. thank
the foundation for the advancement of theoretical physics and mathematics “Basis”.

\nocite{*}

\bibliography{apssamp}

\providecommand{\noopsort}[1]{}\providecommand{\singleletter}[1]{#1}%
\begin{thebibliography}{46}%
\makeatletter
\providecommand \@ifxundefined [1]{%
 \@ifx{#1\undefined}
}%
\providecommand \@ifnum [1]{%
 \ifnum #1\expandafter \@firstoftwo
 \else \expandafter \@secondoftwo
 \fi
}%
\providecommand \@ifx [1]{%
 \ifx #1\expandafter \@firstoftwo
 \else \expandafter \@secondoftwo
 \fi
}%
\providecommand \natexlab [1]{#1}%
\providecommand \enquote  [1]{``#1''}%
\providecommand \bibnamefont  [1]{#1}%
\providecommand \bibfnamefont [1]{#1}%
\providecommand \citenamefont [1]{#1}%
\providecommand \href@noop [0]{\@secondoftwo}%
\providecommand \href [0]{\begingroup \@sanitize@url \@href}%
\providecommand \@href[1]{\@@startlink{#1}\@@href}%
\providecommand \@@href[1]{\endgroup#1\@@endlink}%
\providecommand \@sanitize@url [0]{\catcode `\\12\catcode `\$12\catcode `\&12\catcode `\#12\catcode `\^12\catcode `\_12\catcode `\%12\relax}%
\providecommand \@@startlink[1]{}%
\providecommand \@@endlink[0]{}%
\providecommand \url  [0]{\begingroup\@sanitize@url \@url }%
\providecommand \@url [1]{\endgroup\@href {#1}{\urlprefix }}%
\providecommand \urlprefix  [0]{URL }%
\providecommand \Eprint [0]{\href }%
\providecommand \doibase [0]{https://doi.org/}%
\providecommand \selectlanguage [0]{\@gobble}%
\providecommand \bibinfo  [0]{\@secondoftwo}%
\providecommand \bibfield  [0]{\@secondoftwo}%
\providecommand \translation [1]{[#1]}%
\providecommand \BibitemOpen [0]{}%
\providecommand \bibitemStop [0]{}%
\providecommand \bibitemNoStop [0]{.\EOS\space}%
\providecommand \EOS [0]{\spacefactor3000\relax}%
\providecommand \BibitemShut  [1]{\csname bibitem#1\endcsname}%
\let\auto@bib@innerbib\@empty
\bibitem [{\citenamefont {Higgs}(1964)}]{8}%
  \BibitemOpen
  \bibfield  {author} {\bibinfo {author} {\bibfnamefont {P.~W.}\ \bibnamefont {Higgs}},\ }\bibfield  {title} {\bibinfo {title} {Broken symmetries and the masses of gauge bosons},\ }\href@noop {} {\bibfield  {journal} {\bibinfo  {journal} {Phys. Rev. Lett.}\ }\textbf {\bibinfo {volume} {13}},\ \bibinfo {pages} {508} (\bibinfo {year} {1964})}\BibitemShut {NoStop}%
\bibitem [{\citenamefont {Landau}\ and\ \citenamefont {Lifshitz}(2013)}]{2}%
  \BibitemOpen
  \bibfield  {author} {\bibinfo {author} {\bibfnamefont {L.~D.}\ \bibnamefont {Landau}}\ and\ \bibinfo {author} {\bibfnamefont {E.~M.}\ \bibnamefont {Lifshitz}},\ }\href@noop {} {\emph {\bibinfo {title} {Statistical Physics: Volume 5}}},\ Vol.~\bibinfo {volume} {5}\ (\bibinfo  {publisher} {Elsevier},\ \bibinfo {year} {2013})\BibitemShut {NoStop}%
\bibitem [{\citenamefont {El-Ganainy}\ \emph {et~al.}(2018)\citenamefont {El-Ganainy}, \citenamefont {Makris}, \citenamefont {Khajavikhan}, \citenamefont {Musslimani}, \citenamefont {Rotter},\ and\ \citenamefont {Christodoulides}}]{3}%
  \BibitemOpen
  \bibfield  {author} {\bibinfo {author} {\bibfnamefont {R.}~\bibnamefont {El-Ganainy}}, \bibinfo {author} {\bibfnamefont {K.~G.}\ \bibnamefont {Makris}}, \bibinfo {author} {\bibfnamefont {M.}~\bibnamefont {Khajavikhan}}, \bibinfo {author} {\bibfnamefont {Z.~H.}\ \bibnamefont {Musslimani}}, \bibinfo {author} {\bibfnamefont {S.}~\bibnamefont {Rotter}},\ and\ \bibinfo {author} {\bibfnamefont {D.~N.}\ \bibnamefont {Christodoulides}},\ }\bibfield  {title} {\bibinfo {title} {Non-hermitian physics and pt symmetry},\ }\href@noop {} {\bibfield  {journal} {\bibinfo  {journal} {Nat. Phys.}\ }\textbf {\bibinfo {volume} {14}},\ \bibinfo {pages} {11} (\bibinfo {year} {2018})}\BibitemShut {NoStop}%
\bibitem [{\citenamefont {Makris}\ \emph {et~al.}(2008)\citenamefont {Makris}, \citenamefont {El-Ganainy}, \citenamefont {Christodoulides},\ and\ \citenamefont {Musslimani}}]{7}%
  \BibitemOpen
  \bibfield  {author} {\bibinfo {author} {\bibfnamefont {K.~G.}\ \bibnamefont {Makris}}, \bibinfo {author} {\bibfnamefont {R.}~\bibnamefont {El-Ganainy}}, \bibinfo {author} {\bibfnamefont {D.}~\bibnamefont {Christodoulides}},\ and\ \bibinfo {author} {\bibfnamefont {Z.~H.}\ \bibnamefont {Musslimani}},\ }\bibfield  {title} {\bibinfo {title} {Beam dynamics in p t symmetric optical lattices},\ }\href@noop {} {\bibfield  {journal} {\bibinfo  {journal} {Phys. Rev. Lett.}\ }\textbf {\bibinfo {volume} {100}},\ \bibinfo {pages} {103904} (\bibinfo {year} {2008})}\BibitemShut {NoStop}%
\bibitem [{\citenamefont {Miri}\ and\ \citenamefont {Alu}(2019)}]{6}%
  \BibitemOpen
  \bibfield  {author} {\bibinfo {author} {\bibfnamefont {M.-A.}\ \bibnamefont {Miri}}\ and\ \bibinfo {author} {\bibfnamefont {A.}~\bibnamefont {Alu}},\ }\bibfield  {title} {\bibinfo {title} {Exceptional points in optics and photonics},\ }\href@noop {} {\bibfield  {journal} {\bibinfo  {journal} {Science}\ }\textbf {\bibinfo {volume} {363}},\ \bibinfo {pages} {eaar7709} (\bibinfo {year} {2019})}\BibitemShut {NoStop}%
\bibitem [{\citenamefont {Ozdemir}\ \emph {et~al.}(2019)\citenamefont {Ozdemir}, \citenamefont {Rotter}, \citenamefont {Nori},\ and\ \citenamefont {Yang}}]{5}%
  \BibitemOpen
  \bibfield  {author} {\bibinfo {author} {\bibfnamefont {S.~K.}\ \bibnamefont {Ozdemir}}, \bibinfo {author} {\bibfnamefont {S.}~\bibnamefont {Rotter}}, \bibinfo {author} {\bibfnamefont {F.}~\bibnamefont {Nori}},\ and\ \bibinfo {author} {\bibfnamefont {L.}~\bibnamefont {Yang}},\ }\bibfield  {title} {\bibinfo {title} {Parity–time symmetry and exceptional points in photonics},\ }\href@noop {} {\bibfield  {journal} {\bibinfo  {journal} {Nature Mater.}\ }\textbf {\bibinfo {volume} {18}},\ \bibinfo {pages} {783} (\bibinfo {year} {2019})}\BibitemShut {NoStop}%
\bibitem [{\citenamefont {Bender}\ and\ \citenamefont {Boettcher}(1998)}]{4}%
  \BibitemOpen
  \bibfield  {author} {\bibinfo {author} {\bibfnamefont {C.~M.}\ \bibnamefont {Bender}}\ and\ \bibinfo {author} {\bibfnamefont {S.}~\bibnamefont {Boettcher}},\ }\bibfield  {title} {\bibinfo {title} {Real spectra in non-hermitian hamiltonians having p t symmetry},\ }\href@noop {} {\bibfield  {journal} {\bibinfo  {journal} {Phys. Rev. Lett.}\ }\textbf {\bibinfo {volume} {80}},\ \bibinfo {pages} {5243} (\bibinfo {year} {1998})}\BibitemShut {NoStop}%
\bibitem [{\citenamefont {Feng}\ \emph {et~al.}(2017)\citenamefont {Feng}, \citenamefont {El-Ganainy},\ and\ \citenamefont {Ge}}]{10}%
  \BibitemOpen
  \bibfield  {author} {\bibinfo {author} {\bibfnamefont {L.}~\bibnamefont {Feng}}, \bibinfo {author} {\bibfnamefont {R.}~\bibnamefont {El-Ganainy}},\ and\ \bibinfo {author} {\bibfnamefont {L.}~\bibnamefont {Ge}},\ }\bibfield  {title} {\bibinfo {title} {Non-hermitian photonics based on parity–time symmetry},\ }\href@noop {} {\bibfield  {journal} {\bibinfo  {journal} {Nature Photon.}\ }\textbf {\bibinfo {volume} {11}},\ \bibinfo {pages} {752} (\bibinfo {year} {2017})}\BibitemShut {NoStop}%
\bibitem [{\citenamefont {Peng}\ \emph {et~al.}(2014)\citenamefont {Peng}, \citenamefont {Ozdemir}, \citenamefont {Lei}, \citenamefont {Monifi}, \citenamefont {Gianfreda}, \citenamefont {Long}, \citenamefont {Fan}, \citenamefont {Nori}, \citenamefont {Bender},\ and\ \citenamefont {Yang}}]{11}%
  \BibitemOpen
  \bibfield  {author} {\bibinfo {author} {\bibfnamefont {B.}~\bibnamefont {Peng}}, \bibinfo {author} {\bibfnamefont {S.~K.}\ \bibnamefont {Ozdemir}}, \bibinfo {author} {\bibfnamefont {F.}~\bibnamefont {Lei}}, \bibinfo {author} {\bibfnamefont {F.}~\bibnamefont {Monifi}}, \bibinfo {author} {\bibfnamefont {M.}~\bibnamefont {Gianfreda}}, \bibinfo {author} {\bibfnamefont {G.~L.}\ \bibnamefont {Long}}, \bibinfo {author} {\bibfnamefont {S.}~\bibnamefont {Fan}}, \bibinfo {author} {\bibfnamefont {F.}~\bibnamefont {Nori}}, \bibinfo {author} {\bibfnamefont {C.~M.}\ \bibnamefont {Bender}},\ and\ \bibinfo {author} {\bibfnamefont {L.}~\bibnamefont {Yang}},\ }\bibfield  {title} {\bibinfo {title} {Parity–time-symmetric whispering-gallery microcavities},\ }\href@noop {} {\bibfield  {journal} {\bibinfo  {journal} {Nat. Phys.}\ }\textbf {\bibinfo {volume} {10}},\ \bibinfo {pages} {394} (\bibinfo {year} {2014})}\BibitemShut {NoStop}%
\bibitem [{\citenamefont {Ruter}\ \emph {et~al.}(2010)\citenamefont {Ruter}, \citenamefont {Makris}, \citenamefont {El-Ganainy}, \citenamefont {Christodoulides}, \citenamefont {Segev},\ and\ \citenamefont {Kip}}]{9}%
  \BibitemOpen
  \bibfield  {author} {\bibinfo {author} {\bibfnamefont {C.~E.}\ \bibnamefont {Ruter}}, \bibinfo {author} {\bibfnamefont {K.~G.}\ \bibnamefont {Makris}}, \bibinfo {author} {\bibfnamefont {R.}~\bibnamefont {El-Ganainy}}, \bibinfo {author} {\bibfnamefont {D.~N.}\ \bibnamefont {Christodoulides}}, \bibinfo {author} {\bibfnamefont {M.}~\bibnamefont {Segev}},\ and\ \bibinfo {author} {\bibfnamefont {D.}~\bibnamefont {Kip}},\ }\bibfield  {title} {\bibinfo {title} {Observation of parity–time symmetry in optics},\ }\href@noop {} {\bibfield  {journal} {\bibinfo  {journal} {Nat. Phys.}\ }\textbf {\bibinfo {volume} {6}},\ \bibinfo {pages} {192} (\bibinfo {year} {2010})}\BibitemShut {NoStop}%
\bibitem [{\citenamefont {Zyablovsky}\ \emph {et~al.}(2014)\citenamefont {Zyablovsky}, \citenamefont {Vinogradov}, \citenamefont {Pukhov}, \citenamefont {Dorofeenko},\ and\ \citenamefont {Lisyansky}}]{38}%
  \BibitemOpen
  \bibfield  {author} {\bibinfo {author} {\bibfnamefont {A.~A.}\ \bibnamefont {Zyablovsky}}, \bibinfo {author} {\bibfnamefont {A.~P.}\ \bibnamefont {Vinogradov}}, \bibinfo {author} {\bibfnamefont {A.~A.}\ \bibnamefont {Pukhov}}, \bibinfo {author} {\bibfnamefont {A.~V.}\ \bibnamefont {Dorofeenko}},\ and\ \bibinfo {author} {\bibfnamefont {A.~A.}\ \bibnamefont {Lisyansky}},\ }\bibfield  {title} {\bibinfo {title} {Pt-symmetry in optics},\ }\href@noop {} {\bibfield  {journal} {\bibinfo  {journal} {Phys.-Usp.}\ }\textbf {\bibinfo {volume} {57}},\ \bibinfo {pages} {1063} (\bibinfo {year} {2014})}\BibitemShut {NoStop}%
\bibitem [{\citenamefont {Galda}\ and\ \citenamefont {Vinokur}(2016)}]{12}%
  \BibitemOpen
  \bibfield  {author} {\bibinfo {author} {\bibfnamefont {A.}~\bibnamefont {Galda}}\ and\ \bibinfo {author} {\bibfnamefont {V.~M.}\ \bibnamefont {Vinokur}},\ }\bibfield  {title} {\bibinfo {title} {Parity-time symmetry breaking in magnetic systems},\ }\href@noop {} {\bibfield  {journal} {\bibinfo  {journal} {Phys. Rev. B}\ }\textbf {\bibinfo {volume} {94}},\ \bibinfo {pages} {020408} (\bibinfo {year} {2016})}\BibitemShut {NoStop}%
\bibitem [{\citenamefont {Sadovnikov}\ \emph {et~al.}(2022)\citenamefont {Sadovnikov}, \citenamefont {Zyablovsky}, \citenamefont {Dorofeenko},\ and\ \citenamefont {Nikitov}}]{47}%
  \BibitemOpen
  \bibfield  {author} {\bibinfo {author} {\bibfnamefont {A.~V.}\ \bibnamefont {Sadovnikov}}, \bibinfo {author} {\bibfnamefont {A.~A.}\ \bibnamefont {Zyablovsky}}, \bibinfo {author} {\bibfnamefont {A.~V.}\ \bibnamefont {Dorofeenko}},\ and\ \bibinfo {author} {\bibfnamefont {S.~A.}\ \bibnamefont {Nikitov}},\ }\bibfield  {title} {\bibinfo {title} {Exceptional-point phase transition in coupled magnonic waveguides},\ }\href@noop {} {\bibfield  {journal} {\bibinfo  {journal} {Phys. Rev. Appl.}\ }\textbf {\bibinfo {volume} {18}},\ \bibinfo {pages} {024073} (\bibinfo {year} {2022})}\BibitemShut {NoStop}%
\bibitem [{\citenamefont {Liu}\ \emph {et~al.}(2019)\citenamefont {Liu}, \citenamefont {Sun}, \citenamefont {Zhang}, \citenamefont {Groesbeck}, \citenamefont {Mclaughlin},\ and\ \citenamefont {Vardeny}}]{15}%
  \BibitemOpen
  \bibfield  {author} {\bibinfo {author} {\bibfnamefont {H.}~\bibnamefont {Liu}}, \bibinfo {author} {\bibfnamefont {D.}~\bibnamefont {Sun}}, \bibinfo {author} {\bibfnamefont {C.}~\bibnamefont {Zhang}}, \bibinfo {author} {\bibfnamefont {M.}~\bibnamefont {Groesbeck}}, \bibinfo {author} {\bibfnamefont {R.}~\bibnamefont {Mclaughlin}},\ and\ \bibinfo {author} {\bibfnamefont {Z.~V.}\ \bibnamefont {Vardeny}},\ }\bibfield  {title} {\bibinfo {title} {Observation of exceptional points in magnonic parity-time symmetry devices},\ }\href@noop {} {\bibfield  {journal} {\bibinfo  {journal} {Sci. Adv.}\ }\textbf {\bibinfo {volume} {5}},\ \bibinfo {pages} {eaax9144} (\bibinfo {year} {2019})}\BibitemShut {NoStop}%
\bibitem [{\citenamefont {Temnaya}\ \emph {et~al.}(2022)\citenamefont {Temnaya}, \citenamefont {Safin}, \citenamefont {Kalyabin},\ and\ \citenamefont {Nikitov}}]{18}%
  \BibitemOpen
  \bibfield  {author} {\bibinfo {author} {\bibfnamefont {O.}~\bibnamefont {Temnaya}}, \bibinfo {author} {\bibfnamefont {A.}~\bibnamefont {Safin}}, \bibinfo {author} {\bibfnamefont {D.}~\bibnamefont {Kalyabin}},\ and\ \bibinfo {author} {\bibfnamefont {S.}~\bibnamefont {Nikitov}},\ }\bibfield  {title} {\bibinfo {title} {Parity-time symmetry in planar coupled magnonic heterostructures},\ }\href@noop {} {\bibfield  {journal} {\bibinfo  {journal} {Phys. Rev. Appl.}\ }\textbf {\bibinfo {volume} {18}},\ \bibinfo {pages} {014003} (\bibinfo {year} {2022})}\BibitemShut {NoStop}%
\bibitem [{\citenamefont {Wang}\ \emph {et~al.}(2022)\citenamefont {Wang}, \citenamefont {Jia}, \citenamefont {Lu},\ and\ \citenamefont {Xiong}}]{16}%
  \BibitemOpen
  \bibfield  {author} {\bibinfo {author} {\bibfnamefont {B.}~\bibnamefont {Wang}}, \bibinfo {author} {\bibfnamefont {X.}~\bibnamefont {Jia}}, \bibinfo {author} {\bibfnamefont {X.-H.}\ \bibnamefont {Lu}},\ and\ \bibinfo {author} {\bibfnamefont {H.}~\bibnamefont {Xiong}},\ }\bibfield  {title} {\bibinfo {title} {Pt-symmetric magnon laser in cavity optomagnonics},\ }\href@noop {} {\bibfield  {journal} {\bibinfo  {journal} {Phys. Rev. A}\ }\textbf {\bibinfo {volume} {105}},\ \bibinfo {pages} {053705} (\bibinfo {year} {2022})}\BibitemShut {NoStop}%
\bibitem [{\citenamefont {Wang}\ \emph {et~al.}(2023)\citenamefont {Wang}, \citenamefont {Zeng}, \citenamefont {Guo},\ and\ \citenamefont {Berakdar}}]{17}%
  \BibitemOpen
  \bibfield  {author} {\bibinfo {author} {\bibfnamefont {X.-g.}\ \bibnamefont {Wang}}, \bibinfo {author} {\bibfnamefont {L.-l.}\ \bibnamefont {Zeng}}, \bibinfo {author} {\bibfnamefont {G.-h.}\ \bibnamefont {Guo}},\ and\ \bibinfo {author} {\bibfnamefont {J.}~\bibnamefont {Berakdar}},\ }\bibfield  {title} {\bibinfo {title} {Floquet engineering the exceptional points in parity-time-symmetric magnonics},\ }\href@noop {} {\bibfield  {journal} {\bibinfo  {journal} {Phys. Rev. Lett.}\ }\textbf {\bibinfo {volume} {131}},\ \bibinfo {pages} {186705} (\bibinfo {year} {2023})}\BibitemShut {NoStop}%
\bibitem [{\citenamefont {Zhao}\ \emph {et~al.}(2020)\citenamefont {Zhao}, \citenamefont {Liu}, \citenamefont {Wu}, \citenamefont {Duan}, \citenamefont {Liu},\ and\ \citenamefont {Du}}]{19}%
  \BibitemOpen
  \bibfield  {author} {\bibinfo {author} {\bibfnamefont {J.}~\bibnamefont {Zhao}}, \bibinfo {author} {\bibfnamefont {Y.}~\bibnamefont {Liu}}, \bibinfo {author} {\bibfnamefont {L.}~\bibnamefont {Wu}}, \bibinfo {author} {\bibfnamefont {C.-K.}\ \bibnamefont {Duan}}, \bibinfo {author} {\bibfnamefont {Y.-x.}\ \bibnamefont {Liu}},\ and\ \bibinfo {author} {\bibfnamefont {J.}~\bibnamefont {Du}},\ }\bibfield  {title} {\bibinfo {title} {Observation of anti-pt-symmetry phase transition in the magnon-cavity-magnon coupled system},\ }\href@noop {} {\bibfield  {journal} {\bibinfo  {journal} {Phys. Rev. Appl.}\ }\textbf {\bibinfo {volume} {13}},\ \bibinfo {pages} {014053} (\bibinfo {year} {2020})}\BibitemShut {NoStop}%
\bibitem [{\citenamefont {Djorwe}\ \emph {et~al.}(2019)\citenamefont {Djorwe}, \citenamefont {Pennec},\ and\ \citenamefont {Djafari-Rouhani}}]{20}%
  \BibitemOpen
  \bibfield  {author} {\bibinfo {author} {\bibfnamefont {P.}~\bibnamefont {Djorwe}}, \bibinfo {author} {\bibfnamefont {Y.}~\bibnamefont {Pennec}},\ and\ \bibinfo {author} {\bibfnamefont {B.}~\bibnamefont {Djafari-Rouhani}},\ }\bibfield  {title} {\bibinfo {title} {Exceptional point enhances sensitivity of optomechanical mass sensors},\ }\href@noop {} {\bibfield  {journal} {\bibinfo  {journal} {Phys. Rev. Appl.}\ }\textbf {\bibinfo {volume} {12}},\ \bibinfo {pages} {024002} (\bibinfo {year} {2019})}\BibitemShut {NoStop}%
\bibitem [{\citenamefont {Jing}\ \emph {et~al.}(2017)\citenamefont {Jing}, \citenamefont {Ozdemir}, \citenamefont {Lu},\ and\ \citenamefont {Nori}}]{21}%
  \BibitemOpen
  \bibfield  {author} {\bibinfo {author} {\bibfnamefont {H.}~\bibnamefont {Jing}}, \bibinfo {author} {\bibfnamefont {S.}~\bibnamefont {Ozdemir}}, \bibinfo {author} {\bibfnamefont {H.}~\bibnamefont {Lu}},\ and\ \bibinfo {author} {\bibfnamefont {F.}~\bibnamefont {Nori}},\ }\bibfield  {title} {\bibinfo {title} {High-order exceptional points in optomechanics},\ }\href@noop {} {\bibfield  {journal} {\bibinfo  {journal} {Sci. Rep.}\ }\textbf {\bibinfo {volume} {7}},\ \bibinfo {pages} {1} (\bibinfo {year} {2017})}\BibitemShut {NoStop}%
\bibitem [{\citenamefont {Mukhamedyanov}\ \emph {et~al.}(2023)\citenamefont {Mukhamedyanov}, \citenamefont {Zyablovsky},\ and\ \citenamefont {Andrianov}}]{23}%
  \BibitemOpen
  \bibfield  {author} {\bibinfo {author} {\bibfnamefont {A.}~\bibnamefont {Mukhamedyanov}}, \bibinfo {author} {\bibfnamefont {A.~A.}\ \bibnamefont {Zyablovsky}},\ and\ \bibinfo {author} {\bibfnamefont {E.~S.}\ \bibnamefont {Andrianov}},\ }\bibfield  {title} {\bibinfo {title} {Subthreshold phonon generation in an optomechanical system with an exceptional point},\ }\href@noop {} {\bibfield  {journal} {\bibinfo  {journal} {Opt. Lett.}\ }\textbf {\bibinfo {volume} {48}},\ \bibinfo {pages} {1822} (\bibinfo {year} {2023})}\BibitemShut {NoStop}%
\bibitem [{\citenamefont {Xu}\ \emph {et~al.}(2016)\citenamefont {Xu}, \citenamefont {Mason}, \citenamefont {Jiang},\ and\ \citenamefont {Harris}}]{22}%
  \BibitemOpen
  \bibfield  {author} {\bibinfo {author} {\bibfnamefont {H.}~\bibnamefont {Xu}}, \bibinfo {author} {\bibfnamefont {D.}~\bibnamefont {Mason}}, \bibinfo {author} {\bibfnamefont {L.}~\bibnamefont {Jiang}},\ and\ \bibinfo {author} {\bibfnamefont {J.}~\bibnamefont {Harris}},\ }\bibfield  {title} {\bibinfo {title} {Topological energy transfer in an optomechanical system with exceptional points},\ }\href@noop {} {\bibfield  {journal} {\bibinfo  {journal} {Nature}\ }\textbf {\bibinfo {volume} {537}},\ \bibinfo {pages} {80} (\bibinfo {year} {2016})}\BibitemShut {NoStop}%
\bibitem [{\citenamefont {Jing}\ \emph {et~al.}(2014)\citenamefont {Jing}, \citenamefont {Ozdemir}, \citenamefont {Lu}, \citenamefont {Zhang}, \citenamefont {Yang},\ and\ \citenamefont {Nori}}]{44}%
  \BibitemOpen
  \bibfield  {author} {\bibinfo {author} {\bibfnamefont {H.}~\bibnamefont {Jing}}, \bibinfo {author} {\bibfnamefont {S.}~\bibnamefont {Ozdemir}}, \bibinfo {author} {\bibfnamefont {X.-Y.}\ \bibnamefont {Lu}}, \bibinfo {author} {\bibfnamefont {J.}~\bibnamefont {Zhang}}, \bibinfo {author} {\bibfnamefont {L.}~\bibnamefont {Yang}},\ and\ \bibinfo {author} {\bibfnamefont {F.}~\bibnamefont {Nori}},\ }\bibfield  {title} {\bibinfo {title} {Pt-symmetric phonon laser},\ }\href@noop {} {\bibfield  {journal} {\bibinfo  {journal} {Phys. Rev. Lett.}\ }\textbf {\bibinfo {volume} {113}},\ \bibinfo {pages} {053604} (\bibinfo {year} {2014})}\BibitemShut {NoStop}%
\bibitem [{\citenamefont {Zhang}\ \emph {et~al.}(2015)\citenamefont {Zhang}, \citenamefont {Peng}, \citenamefont {Ozdemir}, \citenamefont {Liu}, \citenamefont {Jing}, \citenamefont {Lu}, \citenamefont {Liu}, \citenamefont {Yang},\ and\ \citenamefont {Nori}}]{45}%
  \BibitemOpen
  \bibfield  {author} {\bibinfo {author} {\bibfnamefont {J.}~\bibnamefont {Zhang}}, \bibinfo {author} {\bibfnamefont {B.}~\bibnamefont {Peng}}, \bibinfo {author} {\bibfnamefont {S.~K.}\ \bibnamefont {Ozdemir}}, \bibinfo {author} {\bibfnamefont {Y.-x.}\ \bibnamefont {Liu}}, \bibinfo {author} {\bibfnamefont {H.}~\bibnamefont {Jing}}, \bibinfo {author} {\bibfnamefont {X.-y.}\ \bibnamefont {Lu}}, \bibinfo {author} {\bibfnamefont {Y.-l.}\ \bibnamefont {Liu}}, \bibinfo {author} {\bibfnamefont {L.}~\bibnamefont {Yang}},\ and\ \bibinfo {author} {\bibfnamefont {F.}~\bibnamefont {Nori}},\ }\bibfield  {title} {\bibinfo {title} {Giant nonlinearity via breaking parity-time symmetry: A route to low-threshold phonon diodes},\ }\href@noop {} {\bibfield  {journal} {\bibinfo  {journal} {Phys. Rev. B}\ }\textbf {\bibinfo {volume} {92}},\ \bibinfo {pages} {115407} (\bibinfo {year} {2015})}\BibitemShut {NoStop}%
\bibitem [{\citenamefont {Lu}\ \emph {et~al.}(2017)\citenamefont {Lu}, \citenamefont {Ozdemir}, \citenamefont {Kuang}, \citenamefont {Nori},\ and\ \citenamefont {Jing}}]{46}%
  \BibitemOpen
  \bibfield  {author} {\bibinfo {author} {\bibfnamefont {H.}~\bibnamefont {Lu}}, \bibinfo {author} {\bibfnamefont {S.}~\bibnamefont {Ozdemir}}, \bibinfo {author} {\bibfnamefont {L.-M.}\ \bibnamefont {Kuang}}, \bibinfo {author} {\bibfnamefont {F.}~\bibnamefont {Nori}},\ and\ \bibinfo {author} {\bibfnamefont {H.}~\bibnamefont {Jing}},\ }\bibfield  {title} {\bibinfo {title} {Exceptional points in random-defect phonon lasers},\ }\href@noop {} {\bibfield  {journal} {\bibinfo  {journal} {Phys. Rev. Appl.}\ }\textbf {\bibinfo {volume} {8}},\ \bibinfo {pages} {044020} (\bibinfo {year} {2017})}\BibitemShut {NoStop}%
\bibitem [{\citenamefont {Brandstetter}\ \emph {et~al.}(2014)\citenamefont {Brandstetter}, \citenamefont {Liertzer}, \citenamefont {Deutsch}, \citenamefont {Klang}, \citenamefont {Schöberl}, \citenamefont {Türeci}, \citenamefont {Strasser}, \citenamefont {Unterrainer},\ and\ \citenamefont {Rotter}}]{24}%
  \BibitemOpen
  \bibfield  {author} {\bibinfo {author} {\bibfnamefont {M.}~\bibnamefont {Brandstetter}}, \bibinfo {author} {\bibfnamefont {M.}~\bibnamefont {Liertzer}}, \bibinfo {author} {\bibfnamefont {C.}~\bibnamefont {Deutsch}}, \bibinfo {author} {\bibfnamefont {P.}~\bibnamefont {Klang}}, \bibinfo {author} {\bibfnamefont {J.}~\bibnamefont {Schöberl}}, \bibinfo {author} {\bibfnamefont {H.~E.}\ \bibnamefont {Türeci}}, \bibinfo {author} {\bibfnamefont {G.}~\bibnamefont {Strasser}}, \bibinfo {author} {\bibfnamefont {K.}~\bibnamefont {Unterrainer}},\ and\ \bibinfo {author} {\bibfnamefont {S.}~\bibnamefont {Rotter}},\ }\bibfield  {title} {\bibinfo {title} {Reversing the pump dependence of a laser at an exceptional point},\ }\href@noop {} {\bibfield  {journal} {\bibinfo  {journal} {Nat. Commun.}\ }\textbf {\bibinfo {volume} {5}},\ \bibinfo {pages} {4034} (\bibinfo {year} {2014})}\BibitemShut {NoStop}%
\bibitem [{\citenamefont {Doronin}\ \emph {et~al.}(2021)\citenamefont {Doronin}, \citenamefont {Zyablovsky},\ and\ \citenamefont {Andrianov}}]{29}%
  \BibitemOpen
  \bibfield  {author} {\bibinfo {author} {\bibfnamefont {I.~V.}\ \bibnamefont {Doronin}}, \bibinfo {author} {\bibfnamefont {A.~A.}\ \bibnamefont {Zyablovsky}},\ and\ \bibinfo {author} {\bibfnamefont {E.~S.}\ \bibnamefont {Andrianov}},\ }\bibfield  {title} {\bibinfo {title} {Strong-coupling-assisted formation of coherent radiation below the lasing threshold},\ }\href@noop {} {\bibfield  {journal} {\bibinfo  {journal} {Opt. Express}\ }\textbf {\bibinfo {volume} {29}},\ \bibinfo {pages} {5624} (\bibinfo {year} {2021})}\BibitemShut {NoStop}%
\bibitem [{\citenamefont {Doronin}\ \emph {et~al.}(2019)\citenamefont {Doronin}, \citenamefont {Zyablovsky}, \citenamefont {Andrianov}, \citenamefont {Pukhov},\ and\ \citenamefont {Vinogradov}}]{27}%
  \BibitemOpen
  \bibfield  {author} {\bibinfo {author} {\bibfnamefont {I.}~\bibnamefont {Doronin}}, \bibinfo {author} {\bibfnamefont {A.}~\bibnamefont {Zyablovsky}}, \bibinfo {author} {\bibfnamefont {E.}~\bibnamefont {Andrianov}}, \bibinfo {author} {\bibfnamefont {A.}~\bibnamefont {Pukhov}},\ and\ \bibinfo {author} {\bibfnamefont {A.}~\bibnamefont {Vinogradov}},\ }\bibfield  {title} {\bibinfo {title} {Lasing without inversion due to parametric instability of the laser near the exceptional point},\ }\href@noop {} {\bibfield  {journal} {\bibinfo  {journal} {Phys. Rev. A}\ }\textbf {\bibinfo {volume} {100}},\ \bibinfo {pages} {021801} (\bibinfo {year} {2019})}\BibitemShut {NoStop}%
\bibitem [{\citenamefont {Feng}\ \emph {et~al.}(2014)\citenamefont {Feng}, \citenamefont {Wong}, \citenamefont {Ma}, \citenamefont {Wang},\ and\ \citenamefont {Zhang}}]{14}%
  \BibitemOpen
  \bibfield  {author} {\bibinfo {author} {\bibfnamefont {L.}~\bibnamefont {Feng}}, \bibinfo {author} {\bibfnamefont {Z.~J.}\ \bibnamefont {Wong}}, \bibinfo {author} {\bibfnamefont {R.-M.}\ \bibnamefont {Ma}}, \bibinfo {author} {\bibfnamefont {Y.}~\bibnamefont {Wang}},\ and\ \bibinfo {author} {\bibfnamefont {X.}~\bibnamefont {Zhang}},\ }\bibfield  {title} {\bibinfo {title} {Single-mode laser by parity-time symmetry breaking},\ }\href@noop {} {\bibfield  {journal} {\bibinfo  {journal} {Science}\ }\textbf {\bibinfo {volume} {346}},\ \bibinfo {pages} {972} (\bibinfo {year} {2014})}\BibitemShut {NoStop}%
\bibitem [{\citenamefont {Hodaei}\ \emph {et~al.}(2014)\citenamefont {Hodaei}, \citenamefont {Miri}, \citenamefont {Heinrich}, \citenamefont {Christodoulides},\ and\ \citenamefont {Khajavikhan}}]{13}%
  \BibitemOpen
  \bibfield  {author} {\bibinfo {author} {\bibfnamefont {H.}~\bibnamefont {Hodaei}}, \bibinfo {author} {\bibfnamefont {M.-A.}\ \bibnamefont {Miri}}, \bibinfo {author} {\bibfnamefont {M.}~\bibnamefont {Heinrich}}, \bibinfo {author} {\bibfnamefont {D.~N.}\ \bibnamefont {Christodoulides}},\ and\ \bibinfo {author} {\bibfnamefont {M.}~\bibnamefont {Khajavikhan}},\ }\bibfield  {title} {\bibinfo {title} {Parity-time–symmetric microring lasers},\ }\href@noop {} {\bibfield  {journal} {\bibinfo  {journal} {Science}\ }\textbf {\bibinfo {volume} {346}},\ \bibinfo {pages} {975} (\bibinfo {year} {2014})}\BibitemShut {NoStop}%
\bibitem [{\citenamefont {Liertzer}\ \emph {et~al.}(2012)\citenamefont {Liertzer}, \citenamefont {Ge}, \citenamefont {Cerjan}, \citenamefont {Stone}, \citenamefont {Türeci},\ and\ \citenamefont {Rotter}}]{28}%
  \BibitemOpen
  \bibfield  {author} {\bibinfo {author} {\bibfnamefont {M.}~\bibnamefont {Liertzer}}, \bibinfo {author} {\bibfnamefont {L.}~\bibnamefont {Ge}}, \bibinfo {author} {\bibfnamefont {A.}~\bibnamefont {Cerjan}}, \bibinfo {author} {\bibfnamefont {A.~D.}\ \bibnamefont {Stone}}, \bibinfo {author} {\bibfnamefont {H.~E.}\ \bibnamefont {Türeci}},\ and\ \bibinfo {author} {\bibfnamefont {S.}~\bibnamefont {Rotter}},\ }\bibfield  {title} {\bibinfo {title} {Pump-induced exceptional points in lasers},\ }\href@noop {} {\bibfield  {journal} {\bibinfo  {journal} {Phys. Rev. Lett.}\ }\textbf {\bibinfo {volume} {108}},\ \bibinfo {pages} {173901} (\bibinfo {year} {2012})}\BibitemShut {NoStop}%
\bibitem [{\citenamefont {Pashkevich}\ \emph {et~al.}(2024)\citenamefont {Pashkevich}, \citenamefont {Doronin}, \citenamefont {Andrianov},\ and\ \citenamefont {Zyablovsky}}]{30}%
  \BibitemOpen
  \bibfield  {author} {\bibinfo {author} {\bibfnamefont {I.}~\bibnamefont {Pashkevich}}, \bibinfo {author} {\bibfnamefont {I.}~\bibnamefont {Doronin}}, \bibinfo {author} {\bibfnamefont {E.}~\bibnamefont {Andrianov}},\ and\ \bibinfo {author} {\bibfnamefont {A.}~\bibnamefont {Zyablovsky}},\ }\bibfield  {title} {\bibinfo {title} {Transition from inhomogeneous to homogeneous broadening at a lasing prethreshold},\ }\href@noop {} {\bibfield  {journal} {\bibinfo  {journal} {Phys. Rev. A}\ }\textbf {\bibinfo {volume} {109}},\ \bibinfo {pages} {033506} (\bibinfo {year} {2024})}\BibitemShut {NoStop}%
\bibitem [{\citenamefont {Zhang}\ \emph {et~al.}(2018)\citenamefont {Zhang}, \citenamefont {Peng}, \citenamefont {Ozdemir}, \citenamefont {Pichler}, \citenamefont {Krimer}, \citenamefont {Zhao}, \citenamefont {Nori}, \citenamefont {Liu}, \citenamefont {Rotter},\ and\ \citenamefont {Yang}}]{25}%
  \BibitemOpen
  \bibfield  {author} {\bibinfo {author} {\bibfnamefont {J.}~\bibnamefont {Zhang}}, \bibinfo {author} {\bibfnamefont {B.}~\bibnamefont {Peng}}, \bibinfo {author} {\bibfnamefont {S.~K.}\ \bibnamefont {Ozdemir}}, \bibinfo {author} {\bibfnamefont {K.}~\bibnamefont {Pichler}}, \bibinfo {author} {\bibfnamefont {D.~O.}\ \bibnamefont {Krimer}}, \bibinfo {author} {\bibfnamefont {G.}~\bibnamefont {Zhao}}, \bibinfo {author} {\bibfnamefont {F.}~\bibnamefont {Nori}}, \bibinfo {author} {\bibfnamefont {Y.-x.}\ \bibnamefont {Liu}}, \bibinfo {author} {\bibfnamefont {S.}~\bibnamefont {Rotter}},\ and\ \bibinfo {author} {\bibfnamefont {L.}~\bibnamefont {Yang}},\ }\bibfield  {title} {\bibinfo {title} {A phonon laser operating at an exceptional point},\ }\href@noop {} {\bibfield  {journal} {\bibinfo  {journal} {Nature Photon.}\ }\textbf {\bibinfo {volume} {12}},\ \bibinfo {pages} {479} (\bibinfo {year} {2018})}\BibitemShut {NoStop}%
\bibitem [{\citenamefont {Zyablovsky}\ \emph {et~al.}(2021)\citenamefont {Zyablovsky}, \citenamefont {Doronin}, \citenamefont {Andrianov}, \citenamefont {Pukhov}, \citenamefont {Lozovik}, \citenamefont {Vinogradov},\ and\ \citenamefont {Lisyansky}}]{26}%
  \BibitemOpen
  \bibfield  {author} {\bibinfo {author} {\bibfnamefont {A.~A.}\ \bibnamefont {Zyablovsky}}, \bibinfo {author} {\bibfnamefont {I.~V.}\ \bibnamefont {Doronin}}, \bibinfo {author} {\bibfnamefont {E.~S.}\ \bibnamefont {Andrianov}}, \bibinfo {author} {\bibfnamefont {A.~A.}\ \bibnamefont {Pukhov}}, \bibinfo {author} {\bibfnamefont {Y.~E.}\ \bibnamefont {Lozovik}}, \bibinfo {author} {\bibfnamefont {A.~P.}\ \bibnamefont {Vinogradov}},\ and\ \bibinfo {author} {\bibfnamefont {A.~A.}\ \bibnamefont {Lisyansky}},\ }\bibfield  {title} {\bibinfo {title} {Exceptional points as lasing prethresholds},\ }\href@noop {} {\bibfield  {journal} {\bibinfo  {journal} {Laser Photonics Rev.}\ }\textbf {\bibinfo {volume} {15}},\ \bibinfo {pages} {2000450} (\bibinfo {year} {2021})}\BibitemShut {NoStop}%
\bibitem [{\citenamefont {Chen}\ \emph {et~al.}(2017)\citenamefont {Chen}, \citenamefont {Kaya~Ozdemir}, \citenamefont {Zhao}, \citenamefont {Wiersig},\ and\ \citenamefont {Yang}}]{34}%
  \BibitemOpen
  \bibfield  {author} {\bibinfo {author} {\bibfnamefont {W.}~\bibnamefont {Chen}}, \bibinfo {author} {\bibfnamefont {S.}~\bibnamefont {Kaya~Ozdemir}}, \bibinfo {author} {\bibfnamefont {G.}~\bibnamefont {Zhao}}, \bibinfo {author} {\bibfnamefont {J.}~\bibnamefont {Wiersig}},\ and\ \bibinfo {author} {\bibfnamefont {L.}~\bibnamefont {Yang}},\ }\bibfield  {title} {\bibinfo {title} {Exceptional points enhance sensing in an optical microcavity},\ }\href@noop {} {\bibfield  {journal} {\bibinfo  {journal} {Nature}\ }\textbf {\bibinfo {volume} {548}},\ \bibinfo {pages} {192} (\bibinfo {year} {2017})}\BibitemShut {NoStop}%
\bibitem [{\citenamefont {Hodaei}\ \emph {et~al.}(2017)\citenamefont {Hodaei}, \citenamefont {Hassan}, \citenamefont {Wittek}, \citenamefont {Garcia-Gracia}, \citenamefont {El-Ganainy}, \citenamefont {Christodoulides},\ and\ \citenamefont {Khajavikhan}}]{31}%
  \BibitemOpen
  \bibfield  {author} {\bibinfo {author} {\bibfnamefont {H.}~\bibnamefont {Hodaei}}, \bibinfo {author} {\bibfnamefont {A.~U.}\ \bibnamefont {Hassan}}, \bibinfo {author} {\bibfnamefont {S.}~\bibnamefont {Wittek}}, \bibinfo {author} {\bibfnamefont {H.}~\bibnamefont {Garcia-Gracia}}, \bibinfo {author} {\bibfnamefont {R.}~\bibnamefont {El-Ganainy}}, \bibinfo {author} {\bibfnamefont {D.~N.}\ \bibnamefont {Christodoulides}},\ and\ \bibinfo {author} {\bibfnamefont {M.}~\bibnamefont {Khajavikhan}},\ }\bibfield  {title} {\bibinfo {title} {Enhanced sensitivity at higher-order exceptional points},\ }\href@noop {} {\bibfield  {journal} {\bibinfo  {journal} {Nature}\ }\textbf {\bibinfo {volume} {548}},\ \bibinfo {pages} {187} (\bibinfo {year} {2017})}\BibitemShut {NoStop}%
\bibitem [{\citenamefont {Wang}\ \emph {et~al.}(2020)\citenamefont {Wang}, \citenamefont {Lai}, \citenamefont {Yuan}, \citenamefont {Suh},\ and\ \citenamefont {Vahala}}]{32}%
  \BibitemOpen
  \bibfield  {author} {\bibinfo {author} {\bibfnamefont {H.}~\bibnamefont {Wang}}, \bibinfo {author} {\bibfnamefont {Y.-H.}\ \bibnamefont {Lai}}, \bibinfo {author} {\bibfnamefont {Z.}~\bibnamefont {Yuan}}, \bibinfo {author} {\bibfnamefont {M.-G.}\ \bibnamefont {Suh}},\ and\ \bibinfo {author} {\bibfnamefont {K.}~\bibnamefont {Vahala}},\ }\bibfield  {title} {\bibinfo {title} {Petermann-factor sensitivity limit near an exceptional point in a brillouin ring laser gyroscope},\ }\href@noop {} {\bibfield  {journal} {\bibinfo  {journal} {Nat. Commun.}\ }\textbf {\bibinfo {volume} {11}},\ \bibinfo {pages} {1610} (\bibinfo {year} {2020})}\BibitemShut {NoStop}%
\bibitem [{\citenamefont {Wiersig}(2016)}]{36}%
  \BibitemOpen
  \bibfield  {author} {\bibinfo {author} {\bibfnamefont {J.}~\bibnamefont {Wiersig}},\ }\bibfield  {title} {\bibinfo {title} {Sensors operating at exceptional points: General theory},\ }\href@noop {} {\bibfield  {journal} {\bibinfo  {journal} {Phys. Rev. A}\ }\textbf {\bibinfo {volume} {93}},\ \bibinfo {pages} {033809} (\bibinfo {year} {2016})}\BibitemShut {NoStop}%
\bibitem [{\citenamefont {Wiersig}(2020)}]{33}%
  \BibitemOpen
  \bibfield  {author} {\bibinfo {author} {\bibfnamefont {J.}~\bibnamefont {Wiersig}},\ }\bibfield  {title} {\bibinfo {title} {Prospects and fundamental limits in exceptional point-based sensing},\ }\href@noop {} {\bibfield  {journal} {\bibinfo  {journal} {Nat. Commun.}\ }\textbf {\bibinfo {volume} {11}},\ \bibinfo {pages} {2454} (\bibinfo {year} {2020})}\BibitemShut {NoStop}%
\bibitem [{\citenamefont {Grudinin}\ \emph {et~al.}(2010)\citenamefont {Grudinin}, \citenamefont {Lee}, \citenamefont {Painter},\ and\ \citenamefont {Vahala}}]{1}%
  \BibitemOpen
  \bibfield  {author} {\bibinfo {author} {\bibfnamefont {I.~S.}\ \bibnamefont {Grudinin}}, \bibinfo {author} {\bibfnamefont {H.}~\bibnamefont {Lee}}, \bibinfo {author} {\bibfnamefont {O.}~\bibnamefont {Painter}},\ and\ \bibinfo {author} {\bibfnamefont {K.~J.}\ \bibnamefont {Vahala}},\ }\bibfield  {title} {\bibinfo {title} {Phonon laser action in a tunable two-level system},\ }\href@noop {} {\bibfield  {journal} {\bibinfo  {journal} {Phys. Rev. Lett.}\ }\textbf {\bibinfo {volume} {104}},\ \bibinfo {pages} {083901} (\bibinfo {year} {2010})}\BibitemShut {NoStop}%
\bibitem [{\citenamefont {Scully}\ and\ \citenamefont {Zubairy}(1997)}]{39}%
  \BibitemOpen
  \bibfield  {author} {\bibinfo {author} {\bibfnamefont {M.~O.}\ \bibnamefont {Scully}}\ and\ \bibinfo {author} {\bibfnamefont {M.~S.}\ \bibnamefont {Zubairy}},\ }\href@noop {} {\emph {\bibinfo {title} {Quantum optics}}}\ (\bibinfo  {publisher} {Cambridge university press},\ \bibinfo {year} {1997})\BibitemShut {NoStop}%
\bibitem [{\citenamefont {Mukhamedyanov}\ \emph {et~al.}(2024)\citenamefont {Mukhamedyanov}, \citenamefont {Zyablovsky},\ and\ \citenamefont {Andrianov}}]{37}%
  \BibitemOpen
  \bibfield  {author} {\bibinfo {author} {\bibfnamefont {A.}~\bibnamefont {Mukhamedyanov}}, \bibinfo {author} {\bibfnamefont {A.~A.}\ \bibnamefont {Zyablovsky}},\ and\ \bibinfo {author} {\bibfnamefont {E.~S.}\ \bibnamefont {Andrianov}},\ }\bibfield  {title} {\bibinfo {title} {Hard excitation mode of a system with optomechanical instability},\ }\href@noop {} {\bibfield  {journal} {\bibinfo  {journal} {Opt. Lett.}\ }\textbf {\bibinfo {volume} {49}},\ \bibinfo {pages} {782} (\bibinfo {year} {2024})}\BibitemShut {NoStop}%
\bibitem [{\citenamefont {Peng}\ \emph {et~al.}(2016)\citenamefont {Peng}, \citenamefont {Cao}, \citenamefont {Shen}, \citenamefont {Qu}, \citenamefont {Wen}, \citenamefont {Jiang},\ and\ \citenamefont {Xiao}}]{40}%
  \BibitemOpen
  \bibfield  {author} {\bibinfo {author} {\bibfnamefont {P.}~\bibnamefont {Peng}}, \bibinfo {author} {\bibfnamefont {W.}~\bibnamefont {Cao}}, \bibinfo {author} {\bibfnamefont {C.}~\bibnamefont {Shen}}, \bibinfo {author} {\bibfnamefont {W.}~\bibnamefont {Qu}}, \bibinfo {author} {\bibfnamefont {J.}~\bibnamefont {Wen}}, \bibinfo {author} {\bibfnamefont {L.}~\bibnamefont {Jiang}},\ and\ \bibinfo {author} {\bibfnamefont {Y.}~\bibnamefont {Xiao}},\ }\bibfield  {title} {\bibinfo {title} {Anti-parity--time symmetry with flying atoms},\ }\href@noop {} {\bibfield  {journal} {\bibinfo  {journal} {Nat. Phys.}\ }\textbf {\bibinfo {volume} {12}},\ \bibinfo {pages} {1139} (\bibinfo {year} {2016})}\BibitemShut {NoStop}%
\bibitem [{\citenamefont {Yang}\ \emph {et~al.}(2017)\citenamefont {Yang}, \citenamefont {Liu},\ and\ \citenamefont {You}}]{41}%
  \BibitemOpen
  \bibfield  {author} {\bibinfo {author} {\bibfnamefont {F.}~\bibnamefont {Yang}}, \bibinfo {author} {\bibfnamefont {Y.-C.}\ \bibnamefont {Liu}},\ and\ \bibinfo {author} {\bibfnamefont {L.}~\bibnamefont {You}},\ }\bibfield  {title} {\bibinfo {title} {Anti-pt symmetry in dissipatively coupled optical systems},\ }\href@noop {} {\bibfield  {journal} {\bibinfo  {journal} {Phys. Rev. A}\ }\textbf {\bibinfo {volume} {96}},\ \bibinfo {pages} {053845} (\bibinfo {year} {2017})}\BibitemShut {NoStop}%
\bibitem [{\citenamefont {Konotop}\ and\ \citenamefont {Zezyulin}(2018)}]{42}%
  \BibitemOpen
  \bibfield  {author} {\bibinfo {author} {\bibfnamefont {V.~V.}\ \bibnamefont {Konotop}}\ and\ \bibinfo {author} {\bibfnamefont {D.~A.}\ \bibnamefont {Zezyulin}},\ }\bibfield  {title} {\bibinfo {title} {Odd-time reversal pt symmetry induced by an anti-pt-symmetric medium},\ }\href@noop {} {\bibfield  {journal} {\bibinfo  {journal} {Phys. Rev. Lett.}\ }\textbf {\bibinfo {volume} {120}},\ \bibinfo {pages} {123902} (\bibinfo {year} {2018})}\BibitemShut {NoStop}%
\bibitem [{\citenamefont {Choi}\ \emph {et~al.}(2018)\citenamefont {Choi}, \citenamefont {Hahn}, \citenamefont {Yoon},\ and\ \citenamefont {Song}}]{43}%
  \BibitemOpen
  \bibfield  {author} {\bibinfo {author} {\bibfnamefont {Y.}~\bibnamefont {Choi}}, \bibinfo {author} {\bibfnamefont {C.}~\bibnamefont {Hahn}}, \bibinfo {author} {\bibfnamefont {J.~W.}\ \bibnamefont {Yoon}},\ and\ \bibinfo {author} {\bibfnamefont {S.~H.}\ \bibnamefont {Song}},\ }\bibfield  {title} {\bibinfo {title} {Observation of an anti-pt-symmetric exceptional point and energy-difference conserving dynamics in electrical circuit resonators},\ }\href@noop {} {\bibfield  {journal} {\bibinfo  {journal} {Nat. Commun.}\ }\textbf {\bibinfo {volume} {9}},\ \bibinfo {pages} {2182} (\bibinfo {year} {2018})}\BibitemShut {NoStop}%
\end{thebibliography}%

\end{document}